# Spatiotemporal evolutions of similariton pulses in multimode fibers with Raman amplification

## Leila Graini[1,2,*] and Bülend ortaç[2,*]

[1]Telecommunications Laboratory, 8 Mai 1945-Guelma University, Guelma 24000, Algeria
[2]National Nanotechnology Research Center and Institute of Materials Science and Nanotechnology, Bilkent University, Ankara 06800, Turkey
*graini.leila@univ-guelma.dz, ortac@unam.bilkent.edu.tr

## Abstract

**This letter is to pave the way towards the demonstration of spatiotemporal similariton pulses evolution in passive multimode fibers with Raman amplification. We present numerically these issues in a graded-index and step-index multimode fibers and provide a first look at the complex spatiotemporal dynamics of similariton pulses. The results showed that the similariton pulses can be generated in both multimode fibers. The temporal and the spectral evolution of the pulses can be characterized as parabolic shapes with linear chirp and kW peak power. By compressed these, high energy femtoseconds pulses can be obtained starting initial picoseconds pulses. Spatial beam profile could be preserved in both multimode fibers under the predominantly excitation of the fundamental mode. Specifically, Raman amplification and similariton pulses generation in graded-index multimode fiber improves the spatial beam cleaning process under the different initial condition.**

## Introduction

In recent years, there has been increased interest in nonlinear wave propagation in multimode optical fibers (MM-Fibers).The traditional view of MM-Fibers as undesirable for nonlinear applications is gradually changing, since MM-Fibers supports a variety of new spatiotemporal phenomena [1].The formation of multimode solitons is limited by different group delays associated with different propagating modes and nonlinear coupling terms [2]. On the other hand, graded-index multimode fibers (GRIN-Fs) reduce the complexity due to the self-imaging effect resulting from the equal spacing of the modal wave with low intermodal dispersion. The theory of multimode light propagation in GRIN-Fs was established in the early 1970s [3], very recently, the studies of the complex nonlinear spatiotemporal effects in these MM-Fibers have been investigated [4]. Spatiotemporal optical solitons [5], supercontinuum generation [6], spatiotemporal instability [7] and self-similar fiber laser [8] are already reported. The above mentioned numerical studies and experimental observations are remarkable examples showing the complex nonlinear pulse dynamics in MM-Fibers. Motivated by these works and the others, one may thus naturally wonder whether similariton pulses may also occur in MM-Fibers.

Similariton pulses are shape-preserving waves in which appear due to the interaction of nonlinearity, normal dispersion and gain in optical fibers. This causes the shape of an arbitrary input pulse to converge to a parabolic pulse shape and evolves in a self-similar manner. Temporal similaritons have been studied extensively in the context of single-mode optical fibers (SMFs) [9-12] and many studies (Raman, Erbium and Ytterbium amplification) have already demonstrated interesting properties of amplifiers similaritons pulses in conventional SMFs. The platform based on MM-Fibers could be a good candidate for the amplification process as well and the question of whether spatiotemporal similariton pulses could be formed inside the MM-Fibers.

In this letter, we propose a new configuration to demonstrate the existence of the spatiotemporal similariton pulses in two different passive MM-Fiber types (GRIN-F and Step-Index Fiber (STEP-F)) with Raman amplification. For our numerical studies, we used the generalized multimode nonlinear Schrödinger equation (GMMNLSE) considering the interaction of nonlinearity, normal dispersions, gain and modal interactions represented by mode coupling. The gain is generated by Raman amplification owing to their advantages; it can take place at any passive fiber types and any wavelength. The characteristics of the similariton pulses generated in both MM-Fibers present similar behavior as obtained in SMFs under the predominantly excitation of the fundamental mode. High energy similariton pulses have temporal and spectral parabolic shapes with linear chirp. The pulses generated in both cases can be also compressible down to femtosecond pulse duration with ~ 100 kW of peak power. However, the spatiotemporal dynamics under different initial modes excitation in both MM-Fibers are very different. New type of beam cleaning process based on similariton pulse generation with Raman amplification in GRIN-F could be observed.

## Numerical Results

To investigate the spatiotemporal evolutions of similariton pulses in MM-Fibers with Raman amplification, we used the platform presented in Fig.1.MM-Fibers used in our numerical studies have a parabolic and step reflective index profiles in the GRIN-F and the STEP-F with identical core radius of 25 µm and index contrast of Δ=0.0068, with NA=0.17, respectively. We considered the first six linearly polarized modes (LP) as shown in Fig.1.

We calculated the dispersions, the nonlinear and the coupling coefficients using codes introduced by Wright et al [13]. The MM-Fibers characterized by the dispersions coefficients $\beta_2$ for the six modes are as follows: 16.76ps$^2$/km ($LP_{01}$), 16.75ps$^2$/km ($LP_{11a}$ and $LP_{11b}$) and 16.74ps$^2$/km ($LP_{21a}$, $LP_{21b}$ and $LP_{02}$) for the GRIN-F and 16.49ps$^2$/km ($LP_{01}$), 16.06ps$^2$/km ($LP_{11a}$ and $LP_{11b}$), 15.50ps$^2$/km ($LP_{21a}$ and $LP_{21b}$) and 15.31 ps$^2$/km ($LP_{02}$) for the STEP-F at 1060 nm. The effective area of modes in the STEP-F tend to be larger than in the GRIN-F($A_{eff} = 1/S^K_{PPPP}$ , where $S^K_{PPPP}$ is the coupling term) [14]. This feature makes both MM-Fibers suitable for Raman amplification since the effective area for the fundamental mode in MM-fibers scales as the core radius. The effective nonlinearity of some propagating modes is also comparable with conventional SMFs [4]. The effective areas and the nonlinear coefficients γ of the GRIN-F and the STEP-F are 155.09 µm$^2$, 1.22 W$^{-1}$km$^{-1}$ and 966.43µm$^2$, 0.2 W$^{-1}$km$^{-1}$, respectively.

To study the spatiotemporal nonlinear dynamics of similariton pulses, we used the GMMNLSE [14], solved by the split-step Fourier method [15]. Gain can be added into the GMMNLSE by an additional term $g_p.A_p$, where $A_p$ is the electric field of mode $p$ and $g_p$ is the small-signal gain for the electric field in mode $p$. For short pulses, the gain coefficient is defined as $g(E_p) = g_0/(1 + E_P/E_{sat})$, where $E_{sat}$ is the energy of saturation. The saturation energy for both MM-Fibers is estimated to be about 1.2 nJ. In our numerical studies, we selected the MM-pump source operating at 1010 nm for Raman gain. Gaussian-shaped input pulse operating at 1060 nm with pulse duration of 2 ps, pulse energy of 0.4 nJ, peak power of 200 W and spectral bandwidth of 0.8 nm is used. In our system, the shift of 13.2 THz (corresponding to 50 nm) between the pump and the signal is selected for optimum Raman amplification in the silica fibers. The gain parameters for Raman amplification are the same for both MM-fibers calculated as $g_0$ = 0.035/m for the pump power of 100 W considering the overfilling nature of the pump ($A_{eff}$ = π.(25µm)$^2$).The fiber length of 130 m is chosen to ensuring a sufficient gain.

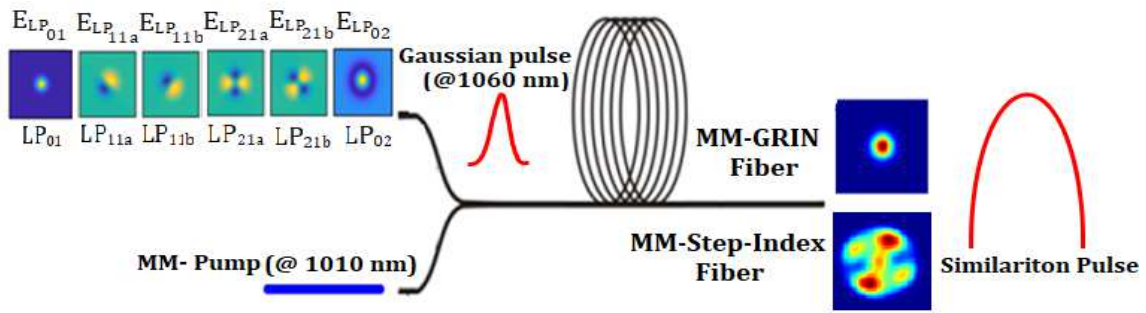

**Fig.1**.Simulationsset-up for spatiotemporal evolutions of similariton pulses in MM-Fibers with Raman amplification.

We first set the initial pulse energy mostly coupled in the $LP_{01}$ mode as 99 %. The temporal and the spectral forms of all the modes for 130 m of pulse propagation in both MM-Fibers can be obtained as shown in Fig.2 (a-d). These results show that only the $LP_{01}$ mode evolves towards a parabolic form and subsequently develops a wide temporal and spectral expansion. As expected for all the other modes, only the $LP_{01}$ mode is strongly amplified and its energy increases exponentially with the fiber length as shown in Fig.3(a and b). It validates the similariton pulses formation of the $LP_{01}$ mode, while the other high-order modes (HOMs) did not converge to the similariton pulses. It can be understood as following: The balance of nonlinearity, dispersion and gain is important for maintaining the parabolic pulse shape during propagation in the fiber. For high initial energy coupling into the $LP_{01}$ mode compared to the other HOMs, the nonlinear phase accumulation is smaller than the dispersion length during propagation. Therefore, for the same amount of gain presented in the STEP-F, the energy can be higher level for the $LP_{01}$ mode with higher nonlinearity. The other HOMs don't transform into the parabolic shape with low initial energy, which is also the result of the inadequate nonlinear phase accumulation.

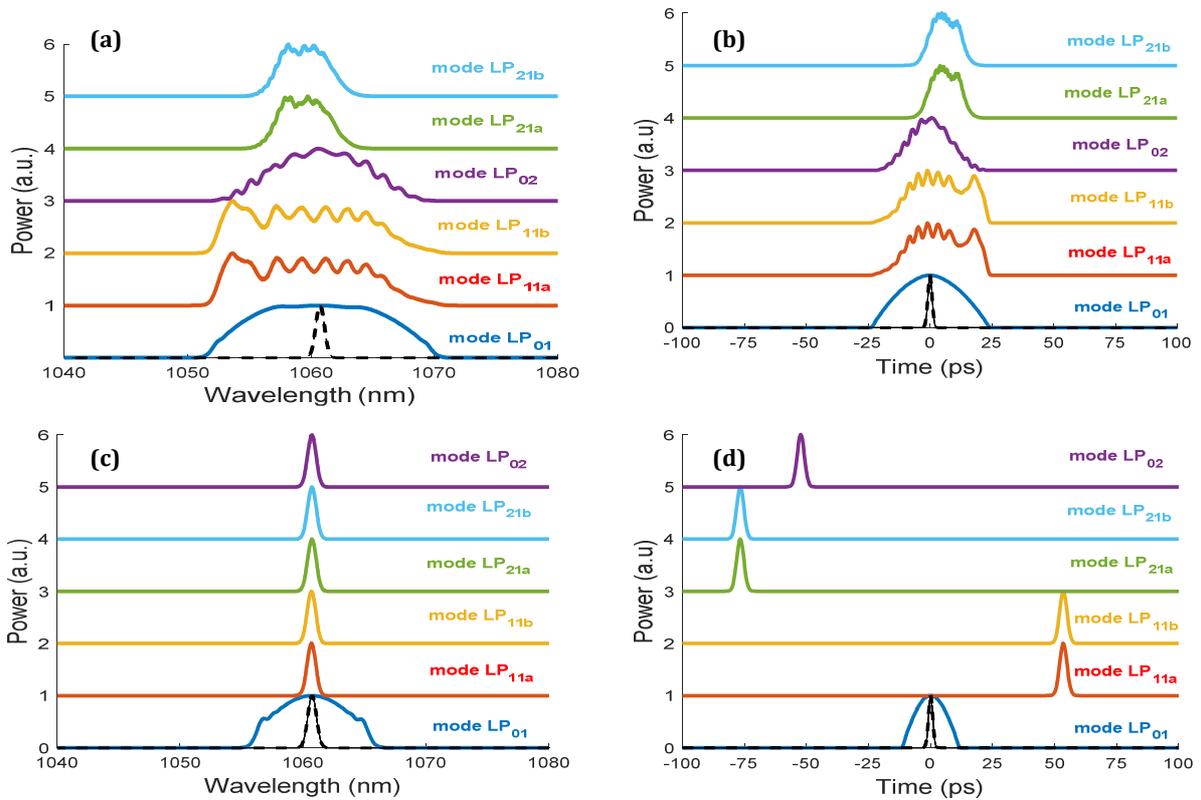

**Fig.2.** Spectral and temporal forms for 130m of pulse propagation for all the modes in the GRIN-F (a and b) and the STEP-F (c and d). Dashed line represents the initial pulse form.

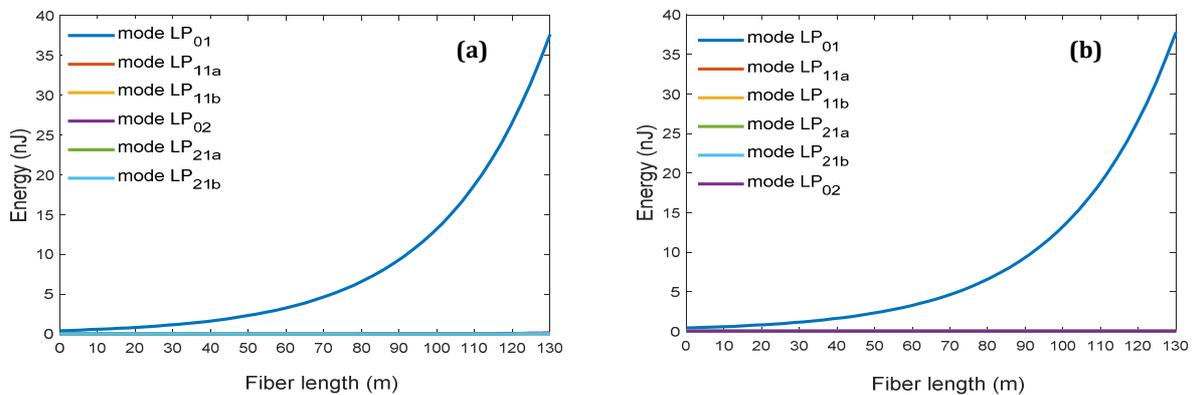

**Fig.3.** Pulse energy evolutions of all the modes in the GRIN-F (a) and the STEP-F (b) where the most initial energy is coupled in the $LP_{01}$ mode.

It should be noticed that the energy exchange is observed between the HOMs in the GRIN-F (Fig. 2.a and b) due to the fact that the intermodal group delays are smaller inside this type of fiber. Therefore, nonlinear coupling among short pulses is maximally achieved along the GRIN-F. On the other hand, the energy coupling to the HOMs was not observed in the STEP-F (Fig. 2.c and d) and each of them evolves as in a single mode case.

Fig.4 shows the formation of similariton pulses of the $LP_{01}$ mode in temporal and spectral domains inside the proposed GRIN-F. Fig.4 (a and c) is highlighting the reshaping of the pulse during the propagation inside the GRIN-F, followed by its self-similar evolution that appeared according to two parts of separate regimes. The first part is corresponding to the restructuring and the reshaping of the $LP_{01}$ mode and the appearance of characteristics of parabolic profile. A typical parabolic shape is obtained at 30 m of pulse propagation inside the GRIN-F, similar to that found in SMFs according to the characteristic length calculated as $Z_c = (3/2g_0)ln\,(Ng_0/6\gamma P_c)$ for N ~ 100[10].In the second part, beyond $Z_C$, the $LP_{01}$ mode has the characteristics of similariton and develops in a self-similar manner with increase in its wide temporal and spectral expansion (Fig.4.a and c). At the output of the GRIN-F, the input Gaussian pulse is now transformed into a parabolic shape with a linear chirp (Fig.4.d). The pulse duration is 32 ps and the pulse energy is about 37.8 nJ. The spectrum generated by the self-similar amplification regime also has a parabolic form (Fig.4.b) with a bandwidth of 16 nm. The small ripples seen in the middle of the spectrum are the result of the limited band with by finite spectral bandwidth of Raman gain (about 23 nm according to the literature [18]. At the same time, Fig. 5 (a-d) shows the self-similar regime and similariton formation for 130 m of pulse propagation inside the STEP-F although the mechanism of the propagations is different. Specifically, the time duration and the energy of the output pulse are 15.5 ps and 37.8 nJ, respectively (Fig.5.d). Here, the input pulse has evolved as a parabolic shape beyond 55 m of propagation which is just a transient state of the pulse evolution in the fiber and under amplification (Fig 5.a and c). The ratio between the nonlinearity coefficient (γ) and the dispersion (β₂) is high in the GRIN-F compared to the STEP-F. Therefore, we observed that the pulse converges to parabolic shape rapidly in the GRIN-F (30 m) than in the STEP-F (55 m). Likewise, the decrease in the nonlinearity coefficient caused less spectral broadening (about 9 nm) for the STEP-F (Fig.5.b).The above results imply that the output pulses have a parabolic pulse shape with high energy amplification. These pulses generated in the GRIN-F and the STEP-F can be compressed down to 255 fs and 340 fs, respectively. ~ 100 kW of peak power and an amplification factor of 500 could be achieved in both cases.

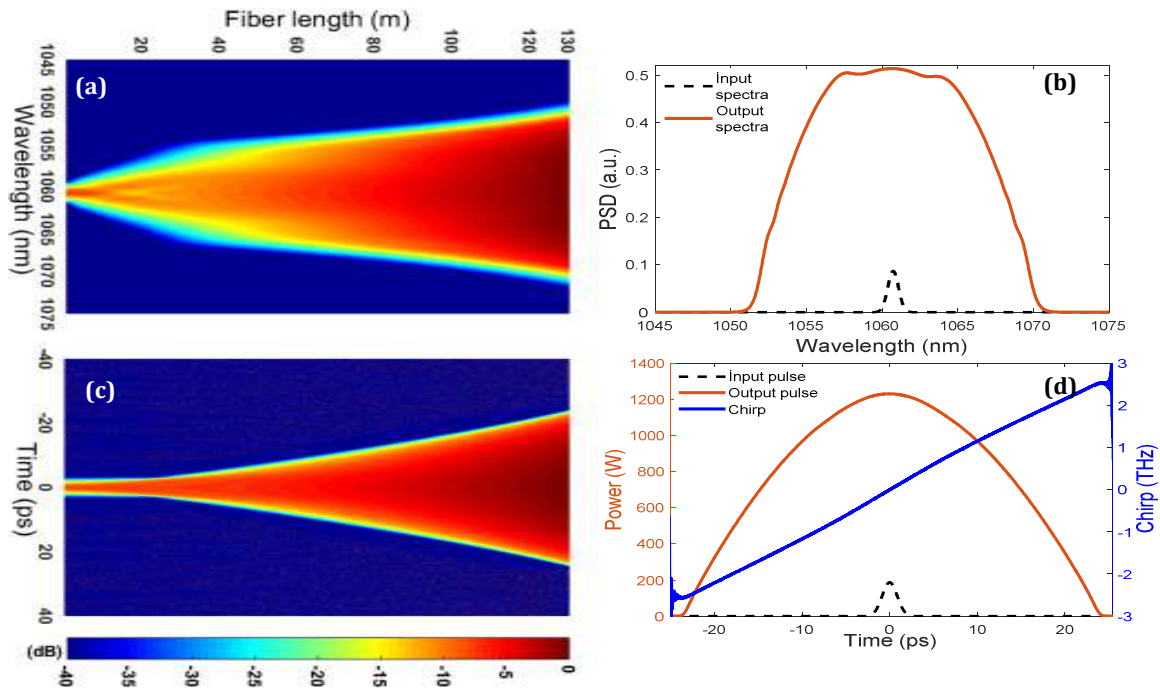

**Fig.4**.The spectral and the temporal similariton evolutions (a and c) and output forms (b and d) of the $LP_{01}$ mode in GRIN-F.

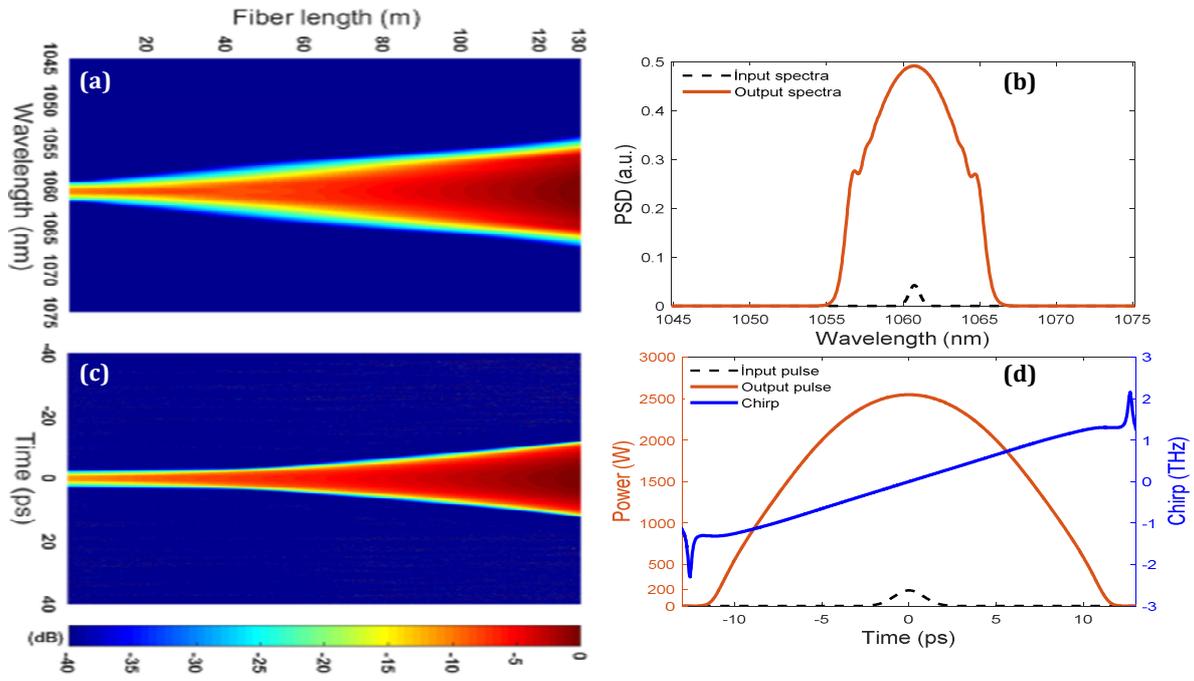

**Fig.5.** The spectral and the temporal similariton evolutions (a and c) and output forms (b and d) of the $LP_{01}$ mode in STEP-F.

Spatial profiles evolution of the total field containing all the modes during the pulse propagation inside the GRIN-F and the STEP-F are presented in Fig.6 (a and b). The excitation of the $LP_{01}$ mode with the highest energy resulting the Gaussian intensity distributions preserved along the both MM-Fibers. Due to the different mode field diameter behavior of $LP_{01}$ mode for both MM-fibers, we also observed the difference beam sizes.

In the second case, we consider the effect of different launch conditions on spatiotemporal pulse evolution. We considered the same initial pulse, GRIN–F and STEP–F parameters. In this part, the simulations are performed to investigate the effect of initial input energy distribution between the modes. To create a significant change compared to previous case, we consider the first six modes with the same initial energy distributions (total peak power of 200 W among all the modes). For same pulse propagation of 130 m fiber length, we noticed that the initial pulse converge to similariton in the GRIN-F for only the $LP_{01}$ mode. The energy of $LP_{01}$ mode is significantly amplified compared to the other HOMs and the other HOMs don't gain energy even they have the same initial energy. It can be explained that the formation of the similariton pulse in the GRIN-F significantly depends on the Raman amplification and mode selective nature that favors the $LP_{01}$ mode [16-17]. On the other hand, in the STEP-F, all the modes evolve towards the respective similariton forms with the same energy due to equal Raman gain. We also observed the energy exchange between the HOMs during the propagation.

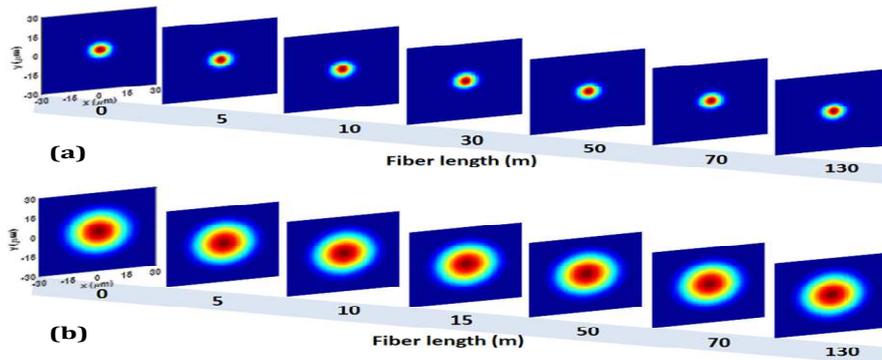

**Fig.6.** Spatial beam evolution in GRIN-F (a) and STEP-F (b) where the most initial energy is coupled in the $LP_{01}$ mode.

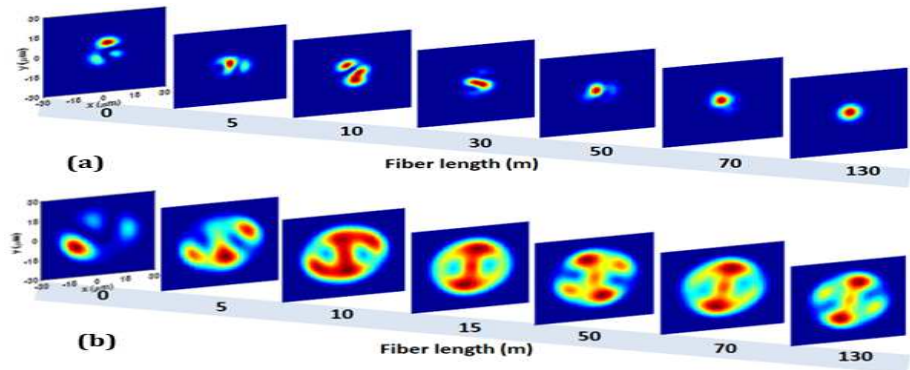

**Fig.7**. Spatial beam evolution in GRIN-F (a) and STEP-F (b) where the initial energy is equally distributed among all the modes.

We also carried out the simulation on the spatial profiles evolution of the total field containing all the modes equally exciting inside the GRIN-F and the STEP-F. Due to the similariton pulse generation occurred only on the $LP_{01}$ mode, an initial spatial beam profile (see Fig.7.a) even composed by the $LP_{01}$ and HOMs is progressively cleaned into a bell-shape beam beyond 50 m of propagation. On the other hand, the spatial beam evolution in the STEP-F presented in Fig.7.b shows that the spatial profile containing all the modes does not converge to a Gaussian like intensity distributions in this case. As a result, the spatiotemporal nonlinear dynamics and the output spatial beam profiles are very different. The similariton generation with the Raman amplification process occurred only on the $LP_{01}$ in the GRIN-F improves the spatial beam cleaning using low energetic input pulses, unlike the cases used rare earth-doped MM-fiber, or used the Kerr effect [19-20].

## Conclusion

In conclusion, we numerically provided a first look at the spatiotemporal similariton pulses evolution in passive MM-Fibers with Raman amplification for the condition of fundamental mode excitation with the most initial energy. We presented these issues in a GRIN-F and STEP-Fusing the same configuration. The similariton pulses have temporal and spectral parabolic shape with linear chirp and high energy in both MM-Fibers although the mechanisms of the propagations between them are different. Compressions of the output amplified pulses leads to the generation of 255 fs and 340 fs pulses in the case of the GRIN-F and the STEP-F, respectively. This approach has a potention in a generation of ultra-short pulses with ~ 100 kW of peak power corresponding to an amplification factor of ~ 500.The transverse distribution intensity of the beams remain Gaussian like profile along the both MM-Fibers. However, the spatiotemporal evolutions as well as the spatial beam profiles are very different in the case of equally distributed initial energy among all the modes. The initial spatial beam profile is progressively cleaned into a bell-shape beam in the GRIN-F because of the selective Raman amplification effect. These features can be useful for new beam cleaning process in the GRIN-F using spatiotemporal similariton evolution. However, the spatial beam profile in the STEP-F contained all the modes with similar energy. Besides, we can also use this platform for the similariton generation for all the modes. From these results, we expect that similariton pulses generation in MM-fibers will open new directions in studies of nonlinear wave propagation and can offer a new route to mode-area scaling for different applications with controllable spatiotemporal properties.